\documentclass[sigconf,screen]{acmart}

\makeatletter
\let\ACM@orig@mkbibcitation\@mkbibcitation
\renewcommand{\@mkbibcitation}{%
  \begingroup
    \emergencystretch=2em
    \hbadness=10000
    \ACM@orig@mkbibcitation
  \endgroup}
\makeatother
\AtBeginDocument{%
  }
    
\usepackage[most]{tcolorbox}
\usepackage{adjustbox}
\usepackage{multirow}
\usepackage{xspace}
\usepackage{algorithm}
\usepackage{algpseudocode}
\usepackage{enumitem}

\newcommand{\dpos}[1]{\textcolor{green!60!black}{#1}}
\newcommand{\dneg}[1]{\textcolor{red!60!black}{#1}}

\setcopyright{cc}
\setcctype{by}
\acmDOI{10.1145/3832783.3834388}
\acmYear{2026}
\copyrightyear{2026}
\acmISBN{979-8-4007-2882-2/2026/10}
\acmConference[ASE '26]{Proceedings of the 41st IEEE/ACM International Conference on Automated Software Engineering}{October 12--16, 2026}{Munich, Germany}
\acmBooktitle{Proceedings of the 41st IEEE/ACM International Conference on Automated Software Engineering (ASE '26), October 12--16, 2026, Munich, Germany}
\acmSubmissionID{ase26main-p1397-p}
\received{2026-03-27}
\received[accepted]{2026-06-18}

\begin{document}



\title{From Runnable Code to Shippable Applications: Test-Driven Development for Full-Stack Web Application Generation}

\author{Yuxuan Wan}
\authornote{Both authors contributed equally to this research.}
\orcid{0009-0006-6739-4675}
\affiliation{%
  \institution{Chinese University of Hong Kong}
  \city{Hong Kong}
  \country{Hong Kong}
}
\email{yxwan@link.cuhk.edu.hk}

\author{Tingshuo Liang}
\authornotemark[1]
\orcid{0009-0002-2022-9497}
\affiliation{%
  \institution{Chinese University of Hong Kong}
  \city{Hong Kong}
  \country{Hong Kong}
}
\email{1155210994@link.cuhk.edu.hk}

\author{Jiakai Xu}
\orcid{0009-0006-2074-9109}
\affiliation{%
  \institution{Columbia University in the City of New York}
  \city{New York}
  \country{USA}
}
\email{ax2155@columbia.edu}

\author{Jingyu Xiao}
\orcid{0000-0002-2394-2995}
\affiliation{%
  \institution{Chinese University of Hong Kong}
  \city{Hong Kong}
  \country{Hong Kong}
}
\email{jyxiao@link.cuhk.edu.hk}

\author{Yintong Huo}
\authornote{Yintong Huo is the corresponding author.}
\orcid{0009-0006-8798-5667}
\affiliation{%
  \institution{Singapore Management University}
  \city{Singapore}
  \country{Singapore}
}
\email{ythuo@smu.edu.sg}

\author{Michael R. Lyu}
\orcid{0000-0002-3666-5798}
\affiliation{%
  \institution{Chinese University of Hong Kong}
  \city{Hong Kong}
  \country{Hong Kong}
}
\email{lyu@cse.cuhk.edu.hk}

\newcommand{\methodname}{TDDev\xspace}
\begin{abstract}\
Coding agents can generate runnable web applications, but their outputs frequently fail to satisfy functional requirements. Although test-driven development (TDD) offers a principled repair loop, applying it to web applications requires deriving executable acceptance tests, validating behavior through dynamic browser interactions, and translating observed failures into actionable feedback. We present \methodname, an experimental instrument that automates these tasks and enables closed-loop TDD with minimal human mediation. Using \methodname, we conduct the first controlled study of TDD for full-stack web application generation across 20 diverse web applications, multiple backbone models, two coding agents, and three TDD implementations. With sufficiently capable backbones, TDD improves accuracy by 15.5--23.7 percentage points and remains effective across both minimal and full-featured coding agents; replacing unreliable feedback with a stronger tester restores positive gains for a lower-capability backbone. Incremental TDD provides bounded control, Whole-Project TDD enables coordinated repair and achieves the highest cost efficiency in most configurations, and Agentic TDD performs best with highly capable models. Based on these findings, we provide a practical decision tree that selects an implementation according to model capability, feedback reliability, and deployment objectives. By automating acceptance testing and failure-guided repair, \methodname also reduces the human testing and intervention required during development.
\end{abstract}



\maketitle

\section{Introduction}
\label{sec:intro}
Web applications are widely used and economically important: reports estimate more than 1.1 billion active websites, with an additional 252,000 new sites launched daily~\cite{website_statistics_2024}. With the development of coding agents~\cite{dong2025survey}, commercial tools already allow users to describe an application and receive a runnable prototype~\cite{lovable}. However, there is a critical distinction between \emph{runnable code} and a \emph{shippable application}: a recent benchmark study shows that applications generated by state-of-the-art agents fail to meet functional requirements in over 70\% of cases~\cite{lu2025webgen}, with users left to manually identify and fix failures.

Test-driven development (TDD) is a software engineering practice where developers iteratively write a test for a specific feature and implement code to satisfy that test~\cite{Mathews2025Test}. TDD provides a principled way to close the runnable/shippable gap:
by specifying executable acceptance tests before any code is written, TDD makes requirements concrete and gives the agent an unambiguous target;
by running those tests against the deployed application, the agent can get structured feedback to refine the code to meet the requirements.
When a test fails, the failure directly identifies what is broken and what the expected behavior should have been, thus turning every defect into an actionable repair signal to improve the application. 

Prior work has shown that TDD-style feedback loops can substantially improve traditional coding agents, from repository-level bug fixing~\cite{yang2024sweagent,Zhang2024AutoCodeRover,xia2025agentless} to multi-agent software workflows~\cite{lin2025soen101} and test-first code generation~\cite{fakhoury2024ticoder,Mathews2025Test,alshahwan2024testgenllm,foster2025ach}. However, these approaches all rely on the same kind of feedback: structured text from compilers, test scripts, or terminals that is directly visible to the agent.

Unfortunately, web application development breaks the feedback loop: the previous agents verify their work by running code and reading the resulting output from the compiler or the terminal, while web applications present three challenges that existing TDD-for-agent approaches cannot handle: 
\begin{itemize}[leftmargin=*]
    \item \textbf{Acceptance-test derivation.} Web app requirements usually arrive as high-level natural language (e.g., ``a shopping website''). To support TDD, these requirements must be transformed into executable browser-based acceptance tests: concrete sequences of navigation, input, and click actions paired with observable expected outcomes that a browser agent can execute and judge.
    \item \textbf{Interactive validation.} Correctness cannot be assessed from source files, compilers, or terminals. The application must be deployed and exercised in the browser through simulated user interactions—such as clicking buttons, submitting forms, and navigating across pages. Nor can this process be scripted in advance, because agent-generated implementations are inherently non-deterministic with various UI structures, element hierarchies, or interaction flows across runs.
    \item \textbf{Failure translation.} Web app failures are experiences rather than explicit logs: mistakes such as broken navigation and missing state updates must be observed in the browser and then translated into precise, actionable feedback that an agent can use for repair. These failures are often contextual and user-facing, making them far harder to capture than standard compiler or runtime errors.
    
\end{itemize}

In current practice, human developers perform all three steps manually: they derive browser-based acceptance tests from application requirements, deploy the app and interact with it to observe what is wrong, and translate those observations into text instructions for the agent. This is not only labor-intensive and frustrating~\cite{Becker2025METR}, but also means the TDD loop cannot be automated, making controlled empirical study of TDD strategies for web application generation infeasible.

To make this study possible, we present \textbf{\methodname}, an experimental instrument that automates acceptance-test derivation, interactive validation, and failure translation. Specifically, \methodname derives structured browser-based acceptance tests from natural-language requirements, deploys generated applications and exercises them through browser-based user interaction simulation, and produces structured failure reports for the coding agent. We validate the reliability of \methodname as an experimental instrument in \textbf{RQ1}, where its browser oracle produces zero false negatives.

With \methodname, we conduct the first controlled study of TDD for full-stack web application generation, organized around five questions:

\begin{itemize}[leftmargin=*]
    \item \textbf{Effectiveness (RQ2):} Is TDD effective for full-stack web application generation? \emph{Yes: with sufficiently capable backbones or reliable feedback, TDD improves accuracy by 15.5--23.7 percentage points and remains effective across coding agents.}
    \item \textbf{Implementation (RQ3):} How do different TDD implementations affect performance? \emph{Incremental TDD provides bounded control, Whole-Project TDD enables coordinated but regression-prone repair, and Agentic TDD performs best with highly capable models.}
    \item \textbf{Cost (RQ4):} What is the performance--cost trade-off of TDD? \emph{Whole-Project TDD provides the highest cost efficiency in most configurations, whereas Incremental and Agentic TDD spend more tokens for stronger control or autonomy.}
    \item \textbf{Recommendation (Discussion~1):} How should practitioners select a TDD implementation? \emph{We provide a decision tree based on model capability, feedback reliability, and the practitioner’s objective.}
    \item \textbf{Practical value (Discussion~2):} Does TDD reduce the human effort required to develop web applications? \emph{Yes: automated acceptance testing and failure-guided repair reduce the need for human testing and intervention during development.}
\end{itemize}
In summary, this paper makes the following contributions:
\begin{itemize}[leftmargin=*]
  \item We present \textbf{\methodname}, a modular experimental instrument that automates acceptance-test derivation, interactive validation, and failure translation, enabling closed-loop TDD with minimal human mediation.
  \item We conduct the first empirical study of TDD for full-stack web application generation, organized around five questions concerning effectiveness, implementation, cost, strategy selection, and practical value.
  \item We release \methodname, all experimental data, and evaluation fixtures to support replication and future research.
\end{itemize}
\section{Background}

\subsection{Task Formulation}

Given a high-level textual requirement $T_0$, a coding agent generates a full-stack web application $\mathit{App} = \mathrm{Agent}(T_0)$. The application is considered correct if it is deployable, renders correctly in a browser, and satisfies an acceptance suite $C(T_0)$ derived from $T_0$. Each element of $C(T_0)$ specifies a user-facing interaction and its expected outcome; the application passes if all elements of $C(T_0)$ are satisfied in the deployed environment.

\subsection{Related Work}

\subsubsection{UI Code Generation}
UI code generation produces front-end code from screenshots or design images, progressing from early CNN-based prototyping~\cite{acsirouglu2019automatic,Cizotto2023WebPF,Moran2018MachineLP,Xu2021Image2e,Chen2018FromUI,nguyen2015reverse,beltramelli2018pix2code,Chen2022CodeGF} to MLLM-based approaches with improved visual fidelity~\cite{Si2024Design2CodeHF,Wan2025DivideAC,Wu2025MLLMBasedUA,Gui2025UICopilotAU,zhou2024bridging,Xiao2024Interaction2CodeHF,Xiao2025DesignBenchAC,Wan2024MRWebAE,le2026uibenchkit,Jiang2025ScreenCoder,Xiao2026VisualRepair,Xiao2026ComUICoder,Xiao2026EfficientUICoder}. These works focus on front-end appearance rather than full-stack functionality; WebGenBench~\cite{lu2025webgen} shows that even state-of-the-art systems frequently fail to satisfy functional requirements, highlighting the gap our work addresses.

\subsubsection{Coding Agents}

Coding agents have shown strong performance on repository-level software engineering tasks, including issue resolution~\cite{yang2024sweagent, Zhang2024AutoCodeRover, ruan2025specrover, xia2025agentless}, program repair~\cite{bouzenia2025repairagent, rondon2025passerine}, build automation~\cite{yu2025cxxcrafter, rua2025buildroid}, and multi-agent development workflows~\cite{lin2025soen101, wang2025openhands}. Empirical analysis of agent trajectories reveals behavioral patterns that distinguish successful from failed executions~\cite{bouzenia2025trajectories}. A key enabler shared across these systems is that the execution environment is directly accessible: the agent can run code, read terminal output, and act on compiler or test feedback in a tight loop. Web application development breaks this assumption --- correctness depends on deployment, browser rendering, and realistic user interaction, none of which is captured by terminal or compiler output alone.

\subsubsection{GUI Testing}
Automated GUI testing has been explored via several paradigms. Record-and-replay methods are easy to use but often fragile and costly to maintain as applications evolve~\cite{Yu2023VisionBasedMA}. Random testing tools such as Monkey~\cite{android_monkey} reduce manual effort, but typically provide limited functional coverage. Model-based testing~\cite{Miguel2016GUIAU,Gu2019PracticalGT} offers more structure by deriving test cases from formal models, yet its effectiveness depends on model quality, requires continuous updates, and often ignores GUI semantics. Learning-based methods~\cite{Lan2024DeeplyRA,Pan2020ReinforcementLB,Li2019HumanoidAD}, commonly based on reinforcement learning, can learn testing policies but usually demand substantial training data and adapt poorly to rapidly changing applications, partly due to limited semantic understanding~\cite{Liu2023MakeLA}. More recently, MLLM-based approaches~\cite{Liu2023MakeLA,Liu2024VisiondrivenAM} have begun to incorporate visual semantics and functional structure, offering a promising direction for GUI testing. These approaches demonstrate the importance of UI-level observation for evaluating user-facing systems. However, they are primarily exploratory and not designed to validate specific functional requirements or return actionable repair feedback within a development loop.

\subsubsection{Test-Driven Development}

Test feedback has been shown to improve code generation across a range of tasks. Wang et al.~\cite{Wang2022TestDriven} use test execution signals during training; AutoCodeRover~\cite{Zhang2024AutoCodeRover} and D4C~\cite{xu2025d4c} use test outcomes for fault localization and patch validation in program repair; and Mathews et al.~\cite{Mathews2025Test} empirically demonstrate TDD benefits when tests are provided alongside natural language prompts. TiCoder~\cite{fakhoury2024ticoder} takes an interactive approach, using the LLM to generate clarifying test cases that the user confirms before code generation, achieving a 46\% absolute improvement in pass@1 with just five interactions. ConTested~\cite{dong2025contested} further exploits inter- and intra-consistency among LLM-generated test suites to select higher-quality code without manual oracles. At industrial scale, Meta's TestGen-LLM~\cite{alshahwan2024testgenllm} automatically improves existing human-written tests using LLMs, with 73\% of recommendations accepted in production, and its successor uses mutation testing to guide targeted test generation~\cite{foster2025ach}. A large-scale study across 37 LLMs and five benchmarks~\cite{shang2025llm4ut} further establishes the landscape of LLM capability for unit test generation. These works share a common assumption: tests either already exist or validation feedback is available directly from the terminal or compiler. For web applications, neither assumption holds. Table~\ref{tab:trad-vs-web} summarizes the differences and maps each to a design decision in \methodname.

\begin{table*}[t]
\centering
\small
\caption{TDD for traditional code tasks vs.\ web applications, and \methodname's corresponding design decisions.}
\label{tab:trad-vs-web}
\vspace{-5pt}
\begin{tabular}{@{}p{2.5cm}p{4.5cm}p{4.5cm}p{4.5cm}@{}}
\toprule
\textbf{Aspect}
  & \textbf{Traditional TDD}
  & \textbf{Web Application TDD}
  & \textbf{\methodname's Approach} \\
\midrule
\textbf{Execution}
  & Local code, directly accessible to the agent
  & Deployed server, rendered and exercised in a browser
  & Automatic framework detection and deployment \\
\addlinespace
\textbf{Requirement}
  & Tests are predefined with clear input–output pairs
  & Tests must be derived from vague natural language requirements
  & Acceptance-test generation via LLM \\
\addlinespace
\textbf{Validation}
  & Terminal output and compiler feedback
  & Browser-based simulation of realistic user interactions
  & LLM agent generating Playwright code on the fly \\
\addlinespace
\textbf{Failure translation}
  & Binary errors and stack traces
  & Contextual, user-facing failures observed in the browser
  & Structured failure report from action trajectory \\
\bottomrule
\end{tabular}
\end{table*}

\begin{figure*}[ht]
    \centering
    \includegraphics[width=.93\linewidth]{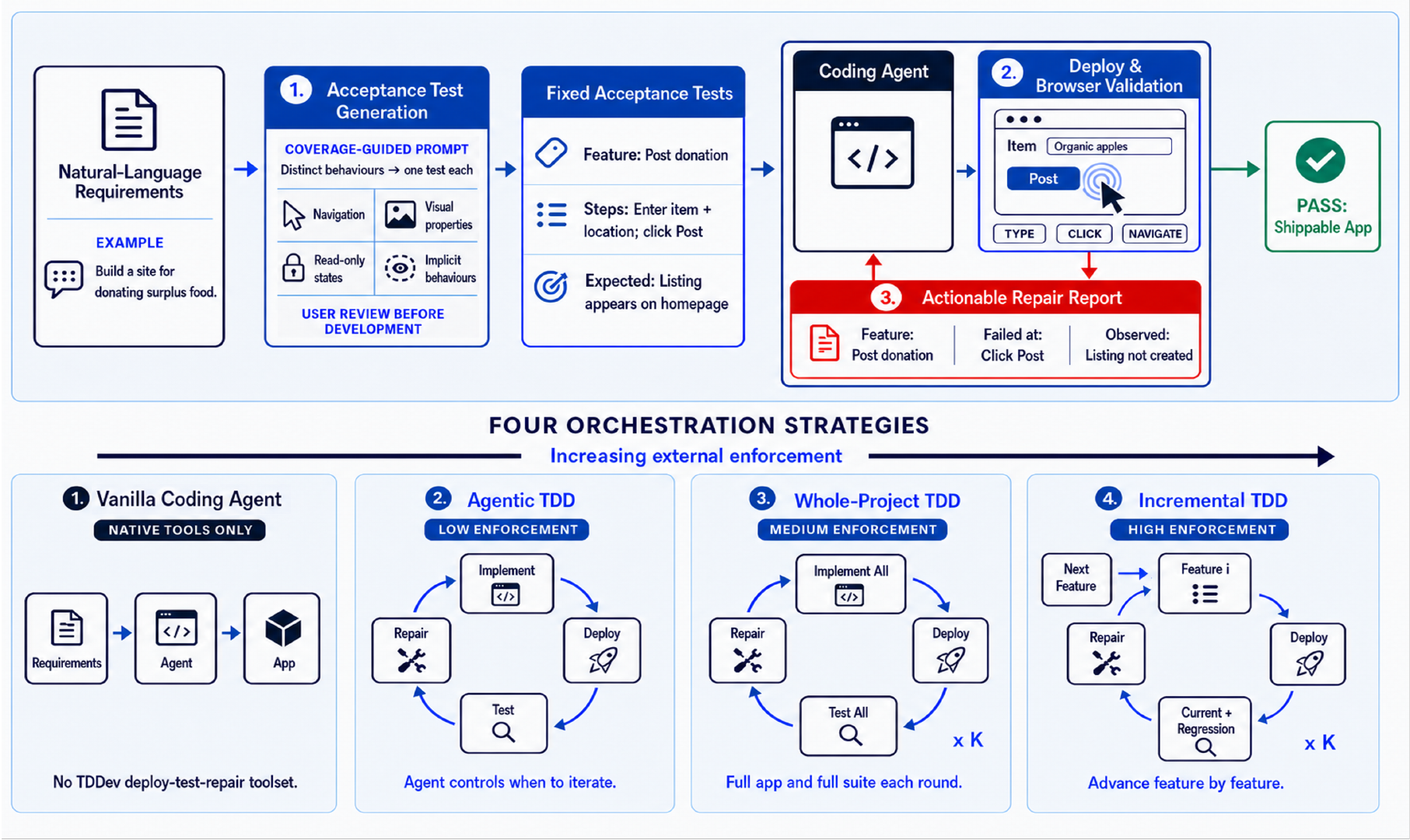}
    \vspace{-5pt}
    \caption{Overview of \methodname. Requirements become user-reviewed
acceptance tests; the coding agent implements the application, which is
deployed and exercised in the browser; failures become structured repair
reports fed back to the agent. Bottom: the same deploy--test--repair tools
are composed into four strategies, ordered by how strongly the harness
enforces their use.}
    \label{fig:method-overview}
\end{figure*}

\section{Methodology}
\label{sec:methodology}

\methodname addresses the three challenges identified in Section~\ref{sec:intro} through a closed test-driven loop with three stages: \textit{acceptance test generation} derives executable tests from natural language requirements before any code is written; \textit{deployment and browser-based validation} deploys the generated application and exercises it through simulated user interactions; and \textit{failure translation} converts browser-observable failures into structured repair reports. Figure~\ref{fig:method-overview} gives an overview. These three stages are composed into four development implementations, which are the experimental variables of our study.

\subsection{Stage 1: Acceptance Test Generation}

This stage derives executable acceptance test cases from natural-language requirements before implementation begins. The tests provide the coding agent with a concrete development target and the repair loop with stable evaluation criteria.

The main challenge is achieving both \textit{validity} (tests grounded in the application's intended behavior), and \textit{coverage} (tests covering distinct behaviors rather than repeated variants of the same interaction). Without guidance, LLMs tend to focus on obvious features, generate overlapping tests, and miss implicit or non-feature requirements.

\methodname uses a structured prompt that asks the model to enumerate the application's distinct behaviors and generate one acceptance test for each. To improve coverage, the prompt explicitly considers navigation between major sections, specified visual properties, read-only browsing states, and behaviors implied but not explicitly stated in the requirement. It also prioritizes breadth over repeated executions of the same feature with different sample data.

Each test is emitted as a structured record containing a \texttt{feature} description (e.g., ``post a product''), an ordered list of interaction \texttt{steps} (e.g., ``enter product name, \ldots, click post''), and an \texttt{expected} outcome observable in the rendered page (e.g., ``the product appears on the homepage''). This representation makes tests executable by the browser agent and provides concrete pass/fail criteria.

The generated tests are presented for user review before development begins. Thus, this stage serves as an authoring aid rather than an autonomous oracle: the user can verify the suite before it enters the development loop, where it serves as the development target.

\subsection{Stage 2: Interactive Validation}
After the coding agent generates the application, this stage verifies whether the implementation satisfies each test case by exercising the app through realistic user interactions.

Web applications must be evaluated in a browser. Scripted tools such as Playwright and Selenium provide precise, reliable interactions, but they assume the app implementation is known in advance. This assumption does not hold for agent-generated applications, whose element structures, labels, and navigation flows may differ across runs. Off-the-shelf GUI agents avoid such assumptions, but they are often imprecise, expensive, and prone to their own errors, which can confound evaluation of the application itself.

To balance reliability and generality, we design a lightweight LLM-backed testing agent. As shown in Algorithm~\ref{alg:test-exec}, before validation, \methodname serves the generated project on a local URL and opens it with Playwright (line 1). At each step, the agent observes the current accessibility tree~\cite{MDNAccessibilityTree}, a structured representation of the rendered page, together with the test context: the feature under test, the interaction steps, the expected outcome, and the trajectory so far. Based on this context, the agent either generates and executes the next Playwright action (line 13) or returns a verdict (``Pass,'' ``Fail,'' or ``Partial'') once it has enough evidence (line 5). After each interaction, the executed action and observed outcome are appended to the trajectory (line 14), enabling the agent to condition subsequent actions and judgments on the full interaction history. Because actions are generated from the page as rendered at runtime, the agent can adapt to different code approaches without prior knowledge of their structure. 

\begin{algorithm}[t]
\caption{Unified Browser Validation Agent}
\label{alg:test-exec}

\begin{algorithmic}[1]
\Require test case $c=\langle \mathit{feature}, \mathit{steps}, \mathit{expected}\rangle$, application URL $u$, LLM $M$, max iterations $T$
\Ensure verdict $v \in \{\texttt{pass}, \texttt{fail}, \texttt{partial}\}$, trajectory $\tau$, failure report $f$
\State $\tau \gets [\,]$; navigate browser to $u$
\For{$t = 1$ \textbf{to} $T$}
    \State $o \gets \textsc{ReadRenderedPage}()$ \Comment{visible text, elements, labels}
    \State $r \gets \textsc{QueryLLM}(M,\; c,\; \tau,\; o)$ \Comment{returns Playwright action or verdict}
    \If{$r$ is a verdict $v$}
        \If{$v = \texttt{pass}$}
            \State \Return $(\texttt{pass},\; \tau,\; \varnothing)$
        \Else
            \State $f \gets \textsc{BuildFailureReport}(c,\; \tau,\; r.\mathit{explanation})$
            \State \Return $(v,\; \tau,\; f)$
        \EndIf
    \EndIf
    \State $\mathit{result} \gets \textsc{ExecutePlaywright}(r)$  \Comment{$r$ is generated Playwright code}
    \State append $(r,\; \mathit{result})$ to $\tau$
\EndFor
\State $f \gets \textsc{BuildFailureReport}(c,\; \tau,\; \text{``max iterations reached''})$
\State \Return $(\texttt{fail},\; \tau,\; f)$
\end{algorithmic}
\end{algorithm}

\subsection{Stage 3: Failure Translation}
A raw browser observation alone is often not meaningful to the coding agent; it becomes actionable only when grounded in the interaction context—what actions were taken, what was observed after each step, and how those observations deviated from the expected outcome. This stage converts the testing agent's interaction trajectory into repair-ready feedback when a test does not pass. Specifically, when the testing agent returns a non-passing verdict, \textsc{BuildFailureReport} summarizes the accumulated trajectory and the agent's natural-language rationale into a structured report that records what was attempted, where the failure occurred, and what was observed.

For example, a failure on a ``user login'' feature may produce:

\begin{tcolorbox}[colback=gray!10, colframe=gray!50, boxrule=0.5pt, arc=2pt, left=6pt, right=6pt, top=2pt, bottom=2pt]

\textit{Feature}: User login.\\
\textit{Failed at}: Submitting the login form---no submit button found.\\
\textit{Observation}: The login form rendered with email and password fields, but no submit control was present.\\
\textit{Steps completed}: Navigated to \texttt{/login}; filled in the email and password fields.
\end{tcolorbox}

This report gives the coding agent a concrete starting point for repair, rather than a vague description of the failure.

\subsection{TDD Implementations}
\label{sec:protocols}

With the TDD infrastructure in place, the degree to which it governs the development process becomes an experimental variable.
The same deploy--test--repair tools can be applied under different levels of enforcement: the system can strictly control when and how they are used, leave the decision to the agent, or not provide them at all.
We define three implementations along this enforcement axis, plus a baseline with no TDD infrastructure.

At the highest level of enforcement is \textbf{Incremental}, which follows TDD discipline most strictly.
The system processes one feature at a time: it first tells the coding agent the overall goal and all acceptance tests, then prompts it to implement the current feature (Line~3 of Algorithm~\ref{alg:incremental}), after which it enters a bounded deploy--test--repair loop (Lines~4--12).
At each attempt, the application is deployed, and the current feature's test is run alongside all previously passing tests as a regression suite (Lines~5--6).
If everything passes or partially passes, the feature is admitted to the regression baseline and the system advances to the next feature (Lines~8--10); otherwise, failures are classified, and the agent is asked to repair (Lines~11--12).
This implementation enforces fine-grained feedback: the agent receives test results for each individual feature before moving on, and regressions in previously passing features are surfaced immediately.

At medium enforcement is \textbf{Whole-Project}.
The agent first implements the entire application in a single pass (Line~1 of Algorithm~\ref{alg:whole}), after which the system enters a bounded deploy--test--repair loop over the full test suite (Lines~2--10).
Each iteration deploys the application, runs all tests, and logs the outcome (Lines~3--5); if all tests pass or partially pass, the loop terminates early (Lines~6--8), otherwise failures are classified, and the agent repairs the whole application at once (Lines~9--10).
Feedback is coarser than in Incremental: the agent sees failures across all features simultaneously, without the incremental anchoring of a regression baseline.

At low enforcement is \textbf{Agentic}.
The agent is given the deploy and test tools and instructed on the TDD workflow, but the system does not enforce any ordering or retry loop.
The agent is invoked once and decides for itself when to deploy, when to run tests, and when to stop.
This condition isolates the effect of workflow knowledge and tool access from the effect of external enforcement.

\textbf{Vanilla coding agent} without the TDDev toolset serves as the baseline.
The agent receives only the requirements and the fixed acceptance test cases, with no TDD tools and no retry loop.
The application is deployed and evaluated once after the agent finishes, representing the current default practice for coding agent--based web development.

All four conditions use the same acceptance tests, the same coding agent, and the same backbone model, isolating the enforcement level as the sole variable.
Comparing Whole-Project, Incremental, and Agentic-TDD against Vanilla Agent measures the overall effect of TDD infrastructure; comparing Whole-Project and Incremental against Agentic-TDD separates external enforcement from agent-driven tool use; and comparing Incremental against Whole-Project isolates the benefit of incremental granularity.

\begin{algorithm}[t]
\caption{Whole-Project Implementation}
\label{alg:whole}

\begin{algorithmic}[1]
\Require test suite $C$, coding agent $\mathcal{A}$, attempt budget $K$
\Ensure application $\mathcal{S}$, logged outcomes $\mathcal{R}$
\State $\mathcal{S} \gets \textsc{ImplementAll}(\mathcal{A}, C)$
\For{$k = 1$ \textbf{to} $K$}
    \State $u \gets \textsc{Deploy}(\mathcal{S})$
    \State $\mathcal{R} \gets \textsc{RunTests}(C, u)$
    \State \textsc{LogOutcome}$(\texttt{whole}, k, \mathcal{R})$
    \If{$\mathcal{R}.\mathit{all\_pass\_or\_partial}$}
        \State \Return $(\mathcal{S}, \mathcal{R})$
    \EndIf
    \State $F \gets \textsc{ReportFailures}(\mathcal{R})$
    \State $\mathcal{S} \gets \textsc{Repair}(\mathcal{A}, \mathcal{S}, F)$
\EndFor
\State \Return $(\mathcal{S}, \mathcal{R})$
\end{algorithmic}
\end{algorithm}

\begin{algorithm}[t]
\caption{Incremental Implementation}
\label{alg:incremental}

\begin{algorithmic}[1]
\Require ordered test cases $C = \langle c_1, \ldots, c_n \rangle$, coding agent $\mathcal{A}$, attempt budget $K$
\Ensure application $\mathcal{S}$, passing regression suite $P$
\State $P \gets [\,]$; $\mathcal{S} \gets \varnothing$  \Comment{empty regression suite, empty application}
\For{each $c_i \in C$}
    \State $\mathcal{S} \gets \textsc{ImplementFeature}(\mathcal{A}, \mathcal{S}, c_i)$
    \For{$k = 1$ \textbf{to} $K$}
        \State $u \gets \textsc{Deploy}(\mathcal{S})$
        \State $\mathcal{R} \gets \textsc{RunTests}(P \cup \{c_i\},\; u)$
        \State \textsc{LogOutcome}$(c_i, k, \mathcal{R})$
        \If{$\mathcal{R}.\mathit{all\_pass\_or\_partial}$}
            \State $P \gets P \cup \{c_i\}$; \textbf{break}
        \EndIf
        \State $F \gets \textsc{ReportFailures}(\mathcal{R})$
        \State $\mathcal{S} \gets \textsc{Repair}(\mathcal{A}, \mathcal{S}, F)$
    \EndFor
\EndFor
\State \Return $(\mathcal{S}, P)$
\end{algorithmic}
\end{algorithm}

\section{Experiment}
\label{sec:experiment}

\subsection{Research Questions}

\begin{itemize}[leftmargin=*]
    \item \textbf{RQ1 (Module Reliability):} How reliable are the individual modules of \methodname? We evaluate test generation coverage against ground-truth requirements and testing agent accuracy against known-correct and known-broken fixture applications.
    \item \textbf{RQ2 (Effectiveness):} Is TDD effective for full-stack web application generation? We compare TDD-enabled coding agents against vanilla coding agents.
    \item \textbf{RQ3 (Implementation):} How do different implementations influence TDD performance? We investigate Whole-Project, Incremental, and Agentic-TDD's effect on TDD performance.
    \item \textbf{RQ4 (Cost):} What is the performance-cost trade-off of TDD? We analyze the token consumption versus performance gain across different implementations.
\end{itemize}

\subsection{Benchmarks}

We randomly sampled 20 web applications from the \textbf{WebGen-Bench}~\cite{lu2025webgen} using a fixed seed (seed=42). The benchmark contains 101 web application-generation tasks with human-validated functional requirements. Each task provides a natural-language instruction describing the application and a list of test cases specifying user-facing functionalities and their expected behaviors.

The sample size was a cost-conscious design choice. Our factorial design includes five agent/model combinations and four protocol conditions, yielding 20 protocol cells evaluated across multiple LLM-driven coding sessions and five deploy--test rounds per case. We used simple random sampling rather than stratified sampling to avoid bias toward easy/hard tasks. Each conclusion reported was evaluated using an appropriate statistical significance test.

\subsection{Coding Agents}

\textbf{Minimal coding agent} is a lightweight coding agent implemented for this study using Anthropic's Claude Agent SDK\footnote{\url{https://docs.anthropic.com/en/docs/build-with-claude/agents}. This is Anthropic's widely adopted framework for building production-ready agentic applications.}. Its purpose is to serve as a minimal baseline so that performance differences across experimental conditions are directly attributable to the TDD infrastructure rather than complex agent sophistication (such as dedicated planning modules, memory, or multi-file context selection).

Operationally, the minimal agent executes a standard agentic loop: given a system prompt, task description, and tool set, the underlying LLM iteratively issues tool calls, which are executed and returned to the model—until it emits a natural language completion. Under baseline conditions, the agent has access to three core tools: \texttt{write\_file}, \texttt{read\_file}, and \texttt{bash}. Under Agentic-TDD, three additional tools are made available:\texttt{start\_app}, \texttt{run\_tests}, and \texttt{stop\_app}, along with system prompt instructions guiding the agent through the full TDD workflow (implement, deploy, test, repair, and repeat).

\textbf{OpenCode}\footnote{\url{https://opencode.ai}} is a fully open-source, terminal-based coding agent that supports any OpenAI-compatible model backend.
It is widely used in the research community as a reproducible, model-agnostic baseline for coding agent studies with 193K GitHub stars.
Under Whole-Project, Incremental, and Vanilla, OpenCode is invoked without modification.
Under Agentic-TDD, \methodname's deploy and test tools are exposed via an MCP server injected into OpenCode's session configuration, giving it the same tool access as the minimal coding agent under Agentic-TDD.

\textbf{MCP integration.}
\methodname's environment-bridging tools are packaged as an MCP (Model Context Protocol) server, exposing \texttt{start\_app}, \texttt{run\_tests}, and \texttt{stop\_app} through a standardized stdio interface.
This makes the tools accessible to any MCP-compatible coding agent without modifying \methodname's internals, and is the mechanism that enables cross-agent evaluation under a consistent tool interface.

For the Incremental and Whole-Project strategies, we cap the feedback loop at \(K=5\) rounds for both coding agents. RQ3 analyzes performance across these rounds.

\subsection{Backbone Models}

We use three backbone models spanning a range of capabilities in our study: \textbf{MiniMax-M3} (open-sourced), \textbf{Claude Sonnet 4.6}, and \textbf{Qwen3.5-397B-A17B} (open-sourced). Table~\ref{tab:backbone-models} reports their Artificial Analysis Agentic and Coding Index scores\footnote{https://artificialanalysis.ai/}, which provide
an external characterization of their relative capabilities. We use all models with temperature 0.0 and reasoning turned off.

\begin{table}[t]
\centering
\caption{Backbone models and their capability-index scores.}
\label{tab:backbone-models}
\begin{tabular}{lcc}
\toprule
\textbf{Model} & \textbf{Agentic Index} & \textbf{Coding Index} \\
\midrule
Claude Sonnet 4.6 & 40.8 & 63.0 \\
MiniMax-M3 & 35.4 & 58.6 \\
Qwen3.5-397B-A17B & 19.8 & 48.2 \\
\bottomrule
\end{tabular}
\end{table}

\begin{table}[t]
    \centering
    \caption{Fixture app examples.}
    \label{tab:fixture}
    \begin{tabular}{ll}
    \toprule
    \textbf{Application} & \textbf{Seeded defect} \\ 
    \midrule 
    \texttt{todo}       & Add button removed from the HTML \\
    \texttt{calculator} & Addition performs subtraction \\
    \texttt{login}      & Login button handler removed \\ \ldots & \emph{(7 more apps)} \\
    \bottomrule
    \end{tabular}
\end{table}

\subsection{Evaluation Metrics}

\subsubsection{RQ1 --- Module reliability.}
\textbf{Test generation coverage} is the fraction of human-authored WebGen-Bench acceptance criteria whose essential actions, details, and outcomes are covered by one or more generated test cases. Matches are determined using the semantic-matching rubric below with Claude-Sonnet-5. Coverage is computed per application and macro-averaged; we also report the generated and ground-truth suite sizes. In the experiment, Claude-Sonnet-4.6 performs the test case generation.

\begin{tcolorbox}[colback=gray!10, colframe=gray!50, boxrule=0.5pt, arc=2pt, left=6pt, right=6pt, top=2pt, bottom=2pt]
\small
\textbf{Semantic-matching prompt (abridged).}
For each reference test, determine whether the generated suite covers all its essential actions, required details, and observable outcomes. Multiple generated tests may jointly cover one reference test; partial overlap or merely referring to the same page, entity, or feature is insufficient. Return one JSON object per reference test containing its ID, the IDs of all generated tests that materially contribute to complete coverage, and a brief justification. 
\end{tcolorbox}

\textbf{Testing agent accuracy} is measured as agreement between the agent's verdicts and predetermined ground-truth verdicts over 40 fixture evaluations: 10 base micro-applications $\times$ 2 variants (correct and broken) $\times$ 2 test cases. Each application has a hand-written correct implementation and a broken twin containing exactly one observable defect. These defects represent failure classes observed in pilot WebGen-Bench runs, including missing UI elements, dead event handlers, and incorrect logic, rather than arbitrary mutations. Table~\ref{tab:fixture} shows three examples of such fixtures. We report accuracy separately for correct and broken variants.

\subsubsection{RQ2--3}

\textbf{Accuracy}
Following WebGen-Bench~\cite{lu2025webgen}, each test case receives a verdict of \textsc{Pass}, \textsc{Fail}, or \textsc{Partial} from the testing agent. Accuracy is computed as:
\begin{equation}
\text{acc} = \frac{N_{\text{Pass}} + 0.5 \times N_{\text{Partial}}}{N_{\text{Total}}} \times 100\%
\end{equation}
The same scoring function is applied at two evaluation tiers. \textbf{Tester accuracy} is measured within the TDD loop on the acceptance-test suite using each model combination's own testing model, whereas \textbf{oracle accuracy}, our headline metric, is measured once after development on the original human-authored WebGen-Bench tests using a fixed, independent model (Claude Sonnet 4.6). For the controlled experiments, we converted these human-authored tests into the acceptance-test format described in Section~3.1, using them as a proxy for user-reviewed acceptance tests supplied to the TDD loop.

\textbf{Significance Test}
To assess whether each TDD strategy improves accuracy, we compare its
per-application accuracy with the corresponding backbone's Vanilla baseline
using a paired one-sided Wilcoxon signed-rank test~\cite{wilcoxon1945individual}.
We test the null hypothesis that TDD provides no improvement
($H_0:\Delta\leq0$) against the alternative that TDD improves accuracy
($H_1:\Delta>0$), where
$\Delta=\mathrm{Acc}_{\mathrm{TDD}}-\mathrm{Acc}_{\mathrm{Vanilla}}$.
Significance is denoted by $^{***}p<.001$, $^{**}p<.01$, $^{*}p<.05$, and
$^{\dagger}p<.1$, with $p<.1$ interpreted as suggestive evidence.
We also report a 95\% percentile-bootstrap CI for the mean paired difference,
estimated using 10,000 resamples~\cite{efron1993introduction}.

\subsubsection{RQ4 --- Cost Efficiency.}
We measure cost efficiency as the oracle-accuracy improvement over the Vanilla baseline per million additional tokens consumed; the metric is defined as:
\begin{equation}
\text{CostEff}
=
\frac{
    \text{Acc}
    -
    \text{Acc}_{\text{Vanilla}}
}{
    \left(
    \text{Tok}
    -
    \text{Tok}_{\text{Vanilla}}
    \right)/10^{6}
}
\quad [\text{pp/Mtoken}],
\end{equation}
where \(\text{Tok}\) is the number of input and output tokens per app. A larger value indicates that the protocol achieves a larger percentage-point improvement for each additional million tokens, and thus higher token efficiency, while a non-positive value indicates no accuracy benefit over Non-TDD despite the additional cost.
\section{Results}
\label{sec:results}

\subsection{RQ1: Module Reliability}

\subsubsection{RQ1.1: Test Generation Coverage}

Table~\ref{tab:rq1-coverage} summarizes test-generation coverage across all 101 WebGen-Bench applications. The test-generation module covers 575 of the 647 reference test cases, corresponding to a pooled coverage of \textbf{88.9\%}. The mean per-application coverage is \textbf{88.7\%} (95\% CI: \textbf{85.6\%--91.4\%}), and \textbf{56 of the 101 applications achieve 100\% coverage}. The generated suites contain an average of 23.5 test cases per application, compared with 6.4 reference test cases, reflecting a finer-grained decomposition of the requirements.

\begin{table}[t]
\centering
\caption{Generated (Gen.) test case coverage (Cvrg.) across 101 applications and 647 ground-truth (GT) test cases.}
\label{tab:rq1-coverage}
\begin{tabular}{lrrrr}
\toprule
\textbf{Cases} & \textbf{\#GT} & \textbf{\#Gen.} &
\textbf{Matched} & \textbf{Cvrg.} \\
\midrule
Full Cvrg. (56/101)    & 360 & 1,317 & 360 & 100.0\% \\
Overall (101/101) &
647 & 2,377 & 575 & 88.9\% \\
Per-app Avg.
    & 6.4 & 23.5 & 5.7 & 88.7\% \\
\bottomrule
\end{tabular}
\end{table}

Of the 72 uncovered reference test cases, 43 concern functional behavior, 23 concern data presentation, and six concern design validation. Many misses involve compound criteria whose specific fields or secondary outcomes are not explicit in the high-level requirement, such as confirmation messages, persisted updates, notifications, ordering constraints, accessibility checks, or responsive behavior across multiple device types. Coverage is highest for design-validation criteria (95.1\%), followed by data-display (87.6\%) and functional criteria (87.3\%).

\subsubsection{RQ1.2: Testing Agent Accuracy}

Table~\ref{tab:rq1-agent} reports per-fixture-app results.
The agent evaluates 20 fixture applications (10 correct, 10 broken), each exercised by 2 test cases, for 40 evaluations.
Overall accuracy is \textbf{87.5\%} (35/40).

\begin{table}[t]
\centering

\caption{Testing agent accuracy summary.}
\label{tab:rq1-agent}
\begin{tabular}{lccc}
\toprule
\textbf{Variant} & \textbf{Evaluations} & \textbf{Correct} & \textbf{Accuracy} \\
\midrule
Correct apps & 20 & 15 & 75.0\% \\
Broken apps  & 20 & 20 & 100.0\% \\
\midrule
\textbf{Overall} & \textbf{40} & \textbf{35} & \textbf{87.5\%} \\
\bottomrule
\end{tabular}
\end{table}

The critical finding is an asymmetry between variants: the agent achieves \textbf{100\% accuracy on broken variants} (20/20 evaluations on broken variants are correct) but 75\% on correct variants (5 false positives).
All 5 failures are false positives on correct applications --- the agent reports failure when the application is functioning correctly.
No false negatives occur: the agent never passes a broken application.

The 5 false positives fall into two categories.
\textbf{Selector generation errors} account for three cases: where the agent generates a Playwright selector that does not match any element and times out.
For example, the calculator's operator button is rendered as \texttt{+} but the test step says ``click the add button''; the agent generates \texttt{text=Add}, which matches nothing.
\textbf{Conservative failures} account for two cases: the agent conservatively rejects even minor differences. For instance,  in a registration form app, the confirmation message uses different wording than the test case's expected string and was rejected by the agent.

This asymmetry is the desirable failure mode for a TDD feedback loop.
A false negative (passing a broken application) would silently propagate defects; this never occurs.
A false positive (failing a correct application) triggers an unnecessary repair round, which is conservative but safe.
The 100\% defect detection rate is the property that matters for the closed-loop system.

\begin{tcolorbox}[colback=gray!10, colframe=gray!50, boxrule=0.5pt, arc=2pt, left=6pt, right=6pt, top=2pt, bottom=2pt]
\textbf{Answer to RQ1.} Test generation module covers 88.9\% of ground-truth features, serving as an effective authoring aid that offers high-coverage test candidates; misses are confined to compound criteria not derivable from high-level requirements. The testing agent catches 100\% of real defects with zero false negatives: residual errors are false positives that trigger unnecessary repair rounds, which is conservative but safe.
\end{tcolorbox}

\subsection{RQ2: TDD Effectiveness}


\begin{figure}[ht]
    \centering
    \includegraphics[width=\linewidth]{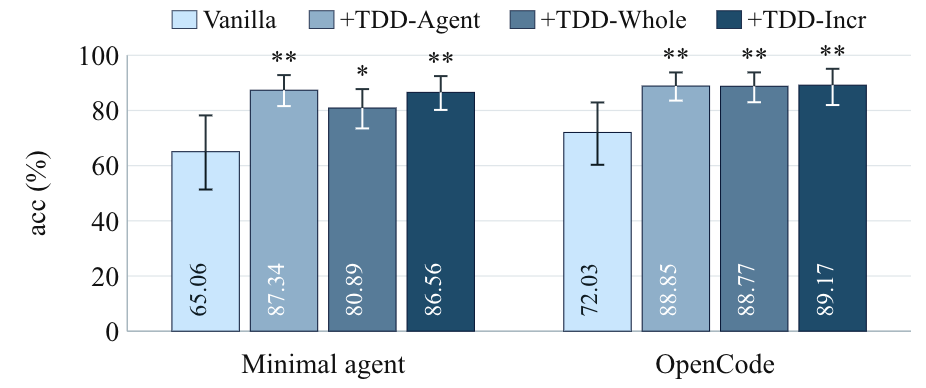}
    \vspace{-15pt}
    \caption{TDD effectiveness across coding agents with MiniMax M3 held fixed. Bars show mean oracle accuracy with 95\% CIs; Significance tests evaluate $H_0:\Delta\leq0$ against $H_1:\Delta>0$ with $^{**}p<.01$, $^{*}p<.05$.}
    \label{fig:rq2-1}
\end{figure}

\subsubsection{Overall Effectiveness}

We first examine whether TDD-based orchestration improves web application generation and whether the improvement persists for an already capable coding agent. Holding the backbone fixed at MiniMax M3, we compare the minimal coding agent, a lightweight API-driven agent limited to basic file and shell operations, with OpenCode, a full-featured coding agent that provides built-in navigation, search, and task-management capabilities. Both agents are evaluated under all four orchestration strategies on the same 20 WebGen-Bench cases. The final application from each successful run is evaluated once by the pinned oracle tester against the original ground-truth test suite. 

Figure~\ref{fig:rq2-1} shows that \emph{every} TDD strategy achieves higher mean oracle accuracy than its corresponding Vanilla baseline for both agents. With the minimal agent, Agentic TDD increases accuracy from 65.1\% to 87.3\% (\(\Delta=22.3\) percentage points), Incremental TDD reaches 86.6\% (\(\Delta=21.5\) percentage points), and Whole-Project TDD reaches 80.9\% (\(\Delta=15.8\) percentage points). OpenCode has a stronger Vanilla baseline of 72.0\%, 7.0 points above that of the minimal agent. Nevertheless, Agentic, Incremental, and Whole-Project TDD reach 88.9\%, 89.2\%, and 88.8\%, respectively, corresponding to gains of 16.7--17.1 pp. (\(p<.01\) for each). 

\subsubsection{Generalization}
We next examine whether the effectiveness of TDD generalizes across model backbones while holding the coding-agent architecture fixed. We evaluate the minimal coding agent with Sonnet and Qwen3.5 as both the developer and test-feedback model. To isolate the role of feedback quality, we additionally evaluate a hybrid configuration in which Qwen3.5 remains the developer while Sonnet provides test feedback. 

\begin{table}[t]
\centering
\caption{Generalization of TDD across backbones with the coding agent fixed to the minimal agent. Acc reports mean oracle accuracy (\%), and $\Delta=\mathrm{Acc}-\mathrm{Acc}_{\mathrm{Vanilla}}$ is measured in percentage points. Significance is denoted by $^{***}p<.001$, $^{**}p<.01$, $^{*}p<.05$, and $^{\dagger}.05\leq p<.10$. The exact $p$-values for the two $\dagger$-marked comparisons are .056 and .092 from top to bottom.}
\label{tab:rq2-2}
\small
\begin{tabular}{lllcc}
\toprule
Backbone & Strategy & Feedback & acc [95\% CI] & $\Delta$\\
\midrule
\multirow{4}{*}{Sonnet}
 & Vanilla     & ---    & 67.8 [58.3, 77.0] & --- \\
 & +TDD-Incr   & Sonnet & 84.6 [77.5, 90.8] & \dpos{16.8$^{***}$} \\
 & +TDD-Whole  & Sonnet & 83.2 [76.1, 90.2] & \dpos{15.5$^{**}$} \\
 & +TDD-Agent  & Sonnet & 91.5 [86.1, 96.5] & \dpos{23.7$^{***}$} \\
\midrule
\multirow{7}{*}{Qwen3.5}
 & Vanilla     & ---    & 75.0 [64.8, 84.5] & --- \\
\cmidrule(l){2-5}
 & \multirow{2}{*}{+TDD-Incr}  & Qwen3.5   & 79.8 [72.9, 86.7] & \dpos{4.8} \\
 &                             & Sonnet & 84.4 [77.0, 91.4] & \dpos{9.4$^{\dagger}$} \\
\cmidrule(l){2-5}
 & \multirow{2}{*}{+TDD-Whole} & Qwen3.5   & 70.6 [62.0, 79.0] & \dneg{$-$4.4} \\
 &                             & Sonnet & 81.8 [75.4, 87.8] & \dpos{6.8$^{\dagger}$} \\
\cmidrule(l){2-5}
 & \multirow{2}{*}{+TDD-Agent} & Qwen3.5   & 74.2 [67.1, 81.5] & \dneg{$-$0.8} \\
 &                             & Sonnet & 84.4 [77.0, 91.2] & \dpos{9.4$^{*}$} \\
\bottomrule
\end{tabular}
\end{table}

\begin{table}[t]
\centering
\caption{Effect of the feedback (fb) model on Qwen3.5's TDD execution traces. Failure-verdict rate is the proportion of test executions labeled \textsc{Fail}; feature-admission and abandonment rates are the proportions of features accepted into or dropped from the application, respectively; and regression rate is the proportion of admitted features that subsequently fail.}
\label{tab:qwen-fb}
\begin{tabular}{lccc}
\toprule
\textbf{Metric} & \textbf{Qwen3.5 fb} &
\textbf{Sonnet fb} & \boldmath$\Delta$ \textbf{(pp)} \\
\midrule
Failure-verdict       & 56.2\% & \textbf{21.3\%} & $-$34.9 \\
Feature-admission    & 51.2\% & \textbf{87.9\%} & $+$36.7 \\
Feature-abandonment   & 48.8\% & \textbf{12.1\%} & $-$36.7 \\
Regression            & 47.6\% & \textbf{14.7\%} & $-$32.9 \\
\bottomrule
\end{tabular}
\end{table}

Table~\ref{tab:rq2-2} shows that TDD produces substantial gains with Sonnet. Relative to its Vanilla accuracy of 67.8\%, Incremental and Whole-Project TDD improve accuracy by 16.8 and 15.5 percentage points, respectively. Agentic TDD achieves the largest improvement, increasing accuracy by 23.7 points to 91.5\%. All three improvements are statistically significant at \(p<.01\).

The results differ when Qwen3.5 serves as the backbone model. Incremental TDD produces a small, non-significant gain of 4.8 points, while Whole-Project and Agentic TDD reduce accuracy by 4.4 and 0.8 points, respectively. TDD therefore does not consistently improve Qwen3.5 when the same backbone must both implement the application and interpret test failures.

Inspection of the development and test-feedback traces explains this result. The Qwen3.5 tester frequently misattributes its own interaction failures to the application: for example, it recognizes that a target exists in the accessibility tree but issues an invalid selector and nevertheless returns \textsc{Fail}. In one case, the same selector error consumes all five feedback rounds; in another, Qwen3.5 repeatedly rejects a color scheme that the final oracle subsequently passes. These erroneous and repetitive verdicts direct the development agent toward unnecessary repairs, cause working regression tests to alternate between \textsc{Pass} and \textsc{Fail}, and exhaust the feedback budget before genuine defects are addressed. 

To isolate the effect of Qwen3.5's unreliable test feedback and test whether the tester is the bottleneck, we introduce a hybrid configuration in which Qwen3.5 remains the developer and only the testing model is replaced with Sonnet. If performance recovers, the result indicates that feedback interpretation, rather than implementation ability, limits Qwen3.5's TDD performance.

Replacing only Qwen3.5's test-feedback model with Sonnet restores positive gains across all three strategies. Incremental and Agentic TDD both reach 84.4\%, while Whole-Project TDD reaches 81.8\%, corresponding to improvements of 6.8--9.4 pp. over Qwen3.5's Vanilla baseline. The replacement also shifts the results from no statistically detectable TDD benefit to suggestive evidence for Incremental and Whole-Project TDD (p = .056 and p = .092) and a statistically significant gain for Agentic TDD (p < .05). The execution traces corroborate this explanation: with Qwen3.5 feedback, 56.2\%  of test executions are labeled as failures, including failures caused by the tester's own interaction errors, compared with 21.3\% under Sonnet feedback (Table~\ref{tab:qwen-fb}). These misleading signals frequently exhaust the five-round repair budget, causing 48.8\% of features to be abandoned; with Sonnet feedback, the abandonment rate falls to 12.1\%.

\begin{tcolorbox}[colback=gray!10, colframe=gray!50, boxrule=0.5pt, arc=2pt, left=6pt, right=6pt, top=2pt, bottom=2pt]
\textbf{Answer to RQ2.}
TDD consistently improves accuracy across coding agents and with capable backbones such as MiniMax M3 and Sonnet. It does not benefit Qwen3.5 with self-generated feedback, but replacing Qwen3.5's tester with Sonnet restores positive gains, identifying feedback quality as the primary bottleneck.
\end{tcolorbox}

\subsection{RQ3: Comparison of Implementations}

\subsubsection{Incremental TDD}

The results reveal a capacity-dependent relationship between the backbone and the preferred TDD implementation. As shown in Table~\ref{tab:backbone-models}, Qwen3.5 has the lowest agentic and coding indices and benefits only from the tightly bounded Incremental strategy (\(+4.8\) points); Whole-Project and Agentic TDD instead reduce its accuracy. MiniMax occupies the middle of the capability range and performs well under both Incremental and Agentic TDD. Agentic produces the largest paired gain with the minimal coding agent (\(+22.3\) points) and Incremental produces the highest final accuracy with OpenCode (89.2\%). Sonnet has the highest capability scores and benefits most from Agentic TDD (\(+23.7\) points), substantially more than from Incremental TDD (\(+16.8\) points). The preferred implementation therefore shifts from bounded incremental control toward autonomous orchestration as model capability increases.

The agent traces support this interpretation. With Qwen3.5 providing its own feedback, only 51.2\% of features are admitted and 47.6\% of admitted features later regress. Less constrained automation amplifies these feedback errors into broader, unnecessary repairs, damaging more established features, which explains why Whole-Project and Agentic TDD fall below Vanilla. Replacing only Qwen3.5's tester with Sonnet raises feature admission to 87.9\%, reduces regression to 14.7\%, and restores positive gains under all three strategies. This shows that Qwen3.5 can implement repairs when given reliable guidance, but its lower feedback-interpretation and agentic capacity prevent it from using autonomous TDD effectively. 

\subsubsection{Whole-Project TDD}

Whole-Project TDD enables coordinated repairs across the application, but its broad edits create greater regression risk than Incremental TDD. Sonnet improves steadily from 61.2\% to 81.7\%, with dips in only 7 of 20 applications. MiniMax peaks at the third attempt and then declines, with regressions in 14 of 20 applications; OpenCode similarly peaks at the fourth evaluation. The traces show that repairs to shared navigation, filtering, styling, and data logic sometimes cause previously passing tests to fail, explaining why Whole-Project TDD underperforms Incremental or Agentic TDD despite improving over Vanilla for capable models.

Qwen3.5's self-judged accuracy increases from 40.6\% to 53.2\%, but its unreliable feedback directs broad, unnecessary repairs, and its final oracle accuracy remains below Vanilla. Replacing only its tester with Sonnet produces a stronger trajectory (73.4\% to 82.5\%), reduces cases with dips from 10 to 8, and changes the final oracle effect from \(-4.4\) to \(+6.8\) points. Whole-Project TDD therefore benefits from capable feedback interpretation, but should retain the best checkpoint because later rounds can regress a stronger intermediate application.

\begin{table}[t]
\centering
\caption{Loop-internal tester accuracy across Whole-Project repair attempts. Compare trajectories within rows, not accuracy levels across rows. Regression (Rgres.) denotes the number of applications whose accuracy declines between consecutive attempts. The first four configurations use the minimal coding agent. Maximum of each row is boldfaced.}
\label{tab:rq3-whole-project}
\small
\begin{tabular}{lrrrrrc}
\toprule
Backbone & $k{=}1$ & $k{=}2$ & $k{=}3$ & $k{=}4$ & $k{=}5$ & Rgres. \\
\midrule
Sonnet 4.6          & 61.2 & 72.4 & 79.4 & 81.3 & \textbf{81.7} & 7/20 \\
MiniMax M3          & 69.4 & 71.8 & \textbf{73.4} & 72.8 & 72.4 & 14/20 \\
Qwen3.5-397B           & 40.6 & 45.1 & 47.9 & 52.6 & \textbf{53.2} & 10/20 \\
Qwen3.5 + Sonnet fb    & 73.4 & 79.9 & 75.9 & 81.2 & \textbf{82.5} & 8/20 \\
\midrule
OpenCode + MiniMax  & 66.8 & 69.0 & 70.4 & \textbf{77.0} & 75.1 & 11/20 \\
\bottomrule
\end{tabular}
\end{table}

\subsubsection{Agentic TDD}

Agentic TDD gives the coding agent control over testing and repair. All agentic sessions invoked \texttt{run\_tests}, averaging 7.3 invocations with Sonnet and 11.8 with MiniMax, yet Sonnet tested less often and achieved the highest accuracy (91.5\%, \(\Delta=23.7\)). Thus, effectiveness depends more on correctly interpreting feedback than on testing frequency. MiniMax also achieves substantial gains with the minimal coding agent (\(\Delta=22.3\)) and OpenCode (\(\Delta=16.8\)).

This autonomy becomes harmful when feedback is unreliable: Qwen3.5 falls below Vanilla (74.2\% vs.\ 75.0\%) and underperforms the harness-controlled Incremental strategy (79.8\%). Replacing its tester with Sonnet restores a 9.4-point gain, confirming that testing and feedback are the bottleneck.

\begin{tcolorbox}[colback=gray!10, colframe=gray!50, boxrule=0.5pt, arc=2pt, left=6pt, right=6pt, top=2pt, bottom=2pt]
\textbf{Answer to RQ3.} The preferred TDD implementation depends on model capability and feedback quality. Incremental TDD provides bounded control for weaker models, Whole-Project TDD enables coordinated repairs but is more prone to regression, and Agentic TDD performs best when the model can interpret feedback and manage the loop reliably.
\end{tcolorbox}

\subsection{RQ4: Cost Efficiency}

We measure cost efficiency as the mean oracle-accuracy gain over Vanilla, in percentage points, divided by the mean additional build-token consumption per application. Build tokens include development and loop testing, while the final oracle evaluation is excluded because it is a measurement cost. Table~\ref{tab:rq4} shows that Whole-Project TDD is the most efficient strategy in four of five configurations, yielding 4.97--10.33 percentage points per million additional tokens. However, it often achieves lower absolute accuracy: with Sonnet, it delivers 65\% of Agentic TDD's gain while using approximately 2.7$\times$ fewer additional tokens. Qwen3.5 with self-feedback is the exception: only Incremental TDD provides a positive return, while Whole-Project and Agentic TDD consume additional tokens without improving accuracy.

\begin{table}[t]
\centering
\caption{Cost efficiency and additional build-token consumption relative to Vanilla. Each cell reports oracle-accuracy gain in percentage points per million additional build tokens, followed by mean additional build tokens per application (M) in parentheses. Build tokens include development and loop testing but exclude final oracle evaluation. The first four configurations use the minimal coding agent.}
\label{tab:rq4}
\small
\begin{tabular}{lccc}
\toprule
\textbf{Backbone} & \textbf{TDD-Incr} &
\textbf{TDD-Whole} & \textbf{TDD-Agent} \\
\midrule
Sonnet 4.6
& 3.94 (4.26)
& \textbf{10.33 (1.50)}
& 5.88 (4.03) \\

MiniMax M3
& 2.85 (7.54)
& \textbf{6.34 (2.49)}
& 2.97 (7.52) \\

Qwen3.5-397B
& \textbf{0.80 (5.99)}
& $-$2.36 (1.86)
& $-$0.23 (3.46) \\

Qwen3.5 + Sonnet fb
& 4.29 (2.19)
& \textbf{5.71 (1.19)}
& 3.50 (2.69) \\
\midrule
OpenCode + MiniMax
& 2.90 (5.90)
& \textbf{4.97 (3.36)}
& 2.33 (7.20) \\
\bottomrule
\end{tabular}
\end{table}

The dominant cost is repeatedly processing context: across all configurations, 96.3\% of build tokens are input and only 3.7\% are output. For configurations using the minimal coding agent, development accounts for 70.1\% of build tokens and testing for 29.9\%. Incremental TDD makes many relatively small calls; for Sonnet, it averages \~285 development calls per application at \~13K tokens per call. Whole-Project TDD occupies the middle in call count (\~82 calls at \~16.2K tokens each), while Agentic TDD makes the fewest but largest calls (\~46 calls at \~72.5K tokens each) because its context grows throughout the session. Whole-Project therefore has the lowest additional cost by amortizing project orientation across features without maintaining Agentic TDD's expanding transcript.

\begin{tcolorbox}[colback=gray!10, colframe=gray!50, boxrule=0.5pt, arc=2pt, left=6pt, right=6pt, top=2pt, bottom=2pt]\
\textbf{Answer to RQ4.} Whole-Project TDD is the most cost-efficient implementation in TDD-effective configurations because it amortizes context processing across the application. Incremental TDD incurs many repeated calls, whereas Agentic TDD makes fewer but substantially larger calls as its context grows.\
\end{tcolorbox}

\section{Discussion}
\label{sec:discussion}

\subsection{Practical Recommendations}

For models with coding capability at least comparable to Qwen3.5-397B-A17B, our results suggest a decision tree based first on feedback reliability and then on model capability and the practitioner's objective:

\begin{itemize}[leftmargin=*]
    \item \textbf{Is the test-feedback model reliable?}
    \begin{itemize}
        \item \textbf{No:} replace it with a stronger tester when possible. Otherwise, use Incremental TDD to bound the effect of erroneous feedback.
        \item \textbf{Yes:} select the implementation according to model capability and objective:
        \begin{itemize}
            \item \textbf{Top-tier model, maximum accuracy:} use Agentic TDD. 
            \item \textbf{Top-tier model, cost efficiency:} use Whole-Project TDD. 
            \item \textbf{Moderate-capacity model, maximum accuracy:} use either Incremental or Agentic TDD.
            \item \textbf{Moderate-capacity model, cost efficiency:} use Whole-Project TDD and retain the best intermediate implementation, rather than the final attempt, to mitigate regression.
            \item \textbf{Lower-capacity model:} use Incremental TDD.
        \end{itemize}
    \end{itemize}
\end{itemize}

Agent sophistication does not remove the need for TDD: all three implementations improve both the minimal coding agent and OpenCode when paired with MiniMax. Instead, implementation choice should match the weakest component of the system, particularly its ability to implement repairs and provide test feedback reliably.

\begin{tcolorbox}[colback=blue!5, colframe=blue!40, boxrule=0.5pt, arc=2pt, left=6pt, right=6pt, top=2pt, bottom=2pt]
\textbf{Takeaway 1.} For models capable of producing adequate implementations, verify feedback quality first. Use Agentic TDD to maximize accuracy with top-tier models, Whole-Project TDD with best-state retention for efficient coordinated repair, and Incremental TDD when model capability or feedback reliability is limited.
\end{tcolorbox}

\begin{table}[ht]
\centering
\caption{Manual intervention comparison between \methodname and Bolt.diy across three developers x 2 sessions each. Manual / Total time is in minutes, Extra Prompt is in number of words.}
\label{tab:user-study}
\begin{tabular}{lccc}
\toprule
\textbf{Method} & \textbf{Manual/Total} & \textbf{Interventions} & \textbf{Extra Prompt} \\
\midrule
Bolt.diy  & 4.7 / 15.2 & 3.0 & 74 \\
+ \methodname & \textbf{0.0} / 18.7 & \textbf{0.0} & \textbf{0} \\
\bottomrule
\end{tabular}
\end{table}
\subsection{Developer Perception}

To complement the automated evaluation, we conducted a user study with three professional developers (two research staff each with at least two prior web application projects, and one front-end developer from a startup) following the methodology of Chen et al.~\cite{Chen2018FromUI}.
Each participant built a web application from a WebGen-Bench requirement twice: once using Bolt.diy + \methodname (MCP tools) and once using vanilla Bolt.diy, an open-source browser-based web generation framework, refining each until reaching a satisfactory state.
We recorded manual intervention time, intervention frequency, and additional prompt length for both tools.

Table~\ref{tab:user-study} shows that \methodname eliminates post-setup intervention.
With Bolt.diy, participants spent an average of 4.7 minutes on manual input across a 15.2-minute session, requiring three rounds of prompting and 74 additional words of guidance.
Critically, this effort was not concentrated at the start: participants had to return to the tool after each generation cycle to test the output, diagnose failures, and formulate corrective instructions.
The agent demanded continuous attention throughout.
With \methodname, participants provided the initial requirement, revised the proposed test cases, and were fully disengaged for the remainder of the session (18.7 minutes on average), with zero additional prompts or interventions required.

The slightly longer total time for \methodname (18.7 vs.\ 15.2 minutes) reflects the cost of automated browser-based testing and iterative repair.
This is not a productivity loss: the extra 3.5 minutes run autonomously while the developer is free to do other work.
Bolt.diy's 15.2 minutes, by contrast, demand active presence for 4.7 of those minutes --- a higher cognitive cost per unit of developer time.

Qualitatively, all three participants described \methodname as fully hands-off and time-saving.
One noted that it ``removes the most frustrating part of the loop --- opening the browser yourself, figuring out what is wrong, and then trying to explain it.''
A second highlighted output quality: ``the app it produces actually works end-to-end, not just visually.''
Participants suggested reducing time spent on non-essential features and adding headless-browser support as future improvements.
These observations are consistent with the quantitative results: the primary value of \methodname lies not in eliminating development time, but in eliminating developer \emph{attention} --- shifting the workload from continuous prompt engineering to autonomous, feedback-driven refinement.

\begin{tcolorbox}[colback=blue!5, colframe=blue!40, boxrule=0.5pt, arc=2pt, left=6pt, right=6pt, top=2pt, bottom=2pt]
\textbf{Takeaway 2.} \methodname eliminates all post-setup intervention (zero prompts, zero redirections), while the baseline requires the developer to return repeatedly throughout the session to diagnose failures and reformulate instructions. \textbf{For agentic tools, the right productivity metric is unattended fraction, not total time.}
\end{tcolorbox}

\subsection{Threats to Validity}

\textbf{Generalizability.}
Our results may not generalize to other models, agent architectures, or application domains. We mitigate this threat by evaluating multiple backbones and coding agents on 20 diverse full-stack applications.

\textbf{Sample size.}
The 20-application sample may underrepresent rare task characteristics and limits statistical power. We therefore report paired significance tests and 95\% confidence intervals for all comparisons to quantify uncertainty rather than relying solely on mean differences.

\textbf{Test-oracle reliability.}
Browser-based testing agents may produce incorrect verdicts because of interaction or interpretation errors. We evaluate oracle reliability separately in RQ1 and use a fixed, independent oracle for final evaluation to avoid condition-specific measurement bias.
\section{Conclusion}
\label{sec:conclusion}

We present \methodname, an experimental instrument that enables closed-loop TDD for full-stack web application generation by automating acceptance-test derivation, dynamic browser validation, and failure translation. Our controlled study shows that TDD substantially improves generation accuracy, but its effectiveness depends on model capability, feedback reliability, and implementation strategy. Incremental TDD provides bounded control, Whole-Project TDD offers coordinated and cost-efficient repair, and Agentic TDD performs best with highly capable models. These findings establish TDD as a practical approach to improving generated web applications while reducing human testing and intervention.

\section*{Data Availability}
All the code and data of \methodname is available at \url{https://github.com/yxwan123/TDD}

\section*{Acknowledgment}
The work was supported by the Research Grants Council of the Hong Kong Special Administrative Region, China, No. SRFS2425-4S03 of the Senior Research Fellow Scheme and Singapore Ministry of Education (MOE) Academic Research Fund (AcRF) Tier 1 grant (Project ID: 23-SIS-SMU-088). Any opinions, findings, conclusions, or recommendations expressed in this material are those of the author(s) and do not reflect the views of the Ministry of Education, Singapore.

\bibliographystyle{ACM-Reference-Format}
\bibliography{sample-base}

\end{document}